\documentclass[dvipdfm]{elsart}

\usepackage{amssymb}    
\usepackage{graphicx}   
\usepackage{hyperref}   

\def\sqrts{$\sqrt{s}$ }
\def\pt{$p_T$ }
\def\mt{$m_T$ }
\def\ee{$e^+e^-$ }
\def\gevc{GeV/$c$}
\def\gev2{GeV/$c^2$}
\def\mevc{MeV/$c$}
\def\mev2{MeV/$c^2$}

\def\pp{p+p }

\begin{document}
\begin{frontmatter}

\title{Dilepton mass spectra in \pp collisions at 
       \sqrts = 200 GeV and the contribution from open charm}

\author{A.~Adare,$^{h}$}
\author{S.~Afanasiev,$^{v}$}
\author{C.~Aidala,$^{i}$}
\author{N.N.~Ajitanand,$^{av}$}
\author{Y.~Akiba,$^{ap,aq}$}
\author{H.~Al-Bataineh,$^{ak}$}
\author{J.~Alexander,$^{av}$}
\author{K.~Aoki,$^{aa,ap}$}
\author{L.~Aphecetche,$^{ay}$}
\author{R.~Armendariz,$^{ak}$}
% ...
\author{(PHENIX Collaboration)}
% ...
\address[h]{University of Colorado, Boulder, CO 80309, USA}
\address[i]{Columbia University, New York, NY 10027 and Nevis Laboratories, Irvington, NY 10533, USA}
% ...
\address[v]{Joint Institute for Nuclear Research, 141980 Dubna, Moscow Region, Russia}
% ...
\address[aa]{Kyoto University, Kyoto 606-8502, Japan}
% ...
\address[ak]{New Mexico State University, Las Cruces, NM 88003, USA}
% ...
\address[ap]{RIKEN, The Institute of Physical and Chemical Research, Wako, Saitama 351-0198, Japan}
\address[aq]{RIKEN BNL Research Center, Brookhaven National Laboratory, Upton, NY 11973-5000, USA}
% ...
\address[av]{Chemistry Department, Stony Brook University, Stony Brook, SUNY, NY 11794-3400, USA}
% ...
\address[ay]{SUBATECH (Ecole des Mines de Nantes, CNRS-IN2P3, Universit{\'e} de Nantes) BP 20722 - 44307, Nantes, France}
% ...
\thanks[deceased]{Deceased}
\thanks[spokes]{PHENIX Spokesperson: jacak@skipper.physics.sunysb.edu}

\date{\today}

\begin{abstract}
PHENIX has measured the electron-positron pair mass spectrum from 0 to
8 \gev2\ in \pp collisions at \sqrts= 200 GeV. The contributions from
light meson decays to \ee pairs have been determined based on
measurements of hadron production cross sections by PHENIX. They
account for nearly all \ee pairs in the mass region below $\sim$1
\gev2. The \ee pair yield remaining after subtracting these
contributions is dominated by semileptonic decays of charmed hadrons
correlated through flavor conservation. Using the spectral shape
predicted by PYTHIA, we estimate the charm production cross section to
be 544 $\pm$ 39(stat) $\pm$ 142(syst) $\pm$ 200(model) $\mu$b, which
is consistent with QCD calculations and measurements of single leptons
by PHENIX.
\end{abstract}

\end{frontmatter}

\maketitle

Because of the large mass of the charm quark, approximately 1.3 \gev2,
it is commonly expected that the charm production cross section can be
calculated in quantum chromo dynamics (QCD) using perturbative methods
(pQCD). Comparing such calculations with experimental data serves as a
test of pQCD and helps to quantify the importance of higher order
terms. Perturbative calculations suggest that charm production at RHIC
energies results primarily from gluon fusion, so charm can probe
gluonic interactions in the matter formed in heavy ion collisions at
RHIC \cite{whitepaper}.  Medium modifications of heavy quark
production and the suppression of bound charmonium states like the
$J/\psi$ have received considerable attention and are thought to be
keys to better understanding properties of strongly interacting
matter. Experiments at RHIC with polarized proton beams will allow the
measurement of spin asymmetries in charm production, which gives
access to the spin contribution of the gluons to the proton in a new
channel \cite{spindocument}.

To date, charm production has been calculated in next-to-leading-order
(NLO) and fixed-order plus next-to-leading-log approximations (FONLL)
\cite{fonll}. These calculations are consistent with the measured D
meson cross sections in 1.96 TeV $p\bar{p}$ collisions published by
CDF \cite{CDF} as well as with single lepton measurements, electrons
\cite{ppg065} and muons \cite{singlemu}, in 200 GeV \pp collisions
from PHENIX. However, the theoretical uncertainties are considerable,
at least a factor of two \cite{fonll} or even larger \cite{vogt}, and
the data prefer larger cross sections within these
uncertainties\footnote{The STAR collaboration reports an even larger
  cross section \cite{star_e,star_d}, which is about a factor of 2-3
  above of what can be accommodated in pQCD calculations.}.  In this
Letter we present a different method to determine the charm cross
section using electron-positron pairs measured with PHENIX during the
RHIC p+p run in 2005.

Electrons are measured in the two PHENIX central arm spectrometers
\cite{phenix}, which each cover $\left|\eta\right|\le0.35$ in
pseudo-rapidity and $\Delta\phi =\pi/2$ in azimuth in a nearly
back-to-back configuration. For charged particles drift chambers (DC)
measure the deflection angles in a magnetic field to determine their
momenta. Ring imaging Cerenkov counters (RICH) as well as
electromagnetic calorimeters (EMCal) distinguish electrons from other
particles. The electron analysis is described in detail in
\cite{ppg065}.

Two data sets are used in the analysis. A reference sample of events
was selected with a minimum bias interaction trigger (MB) that was
based on beam-beam counters (BBC). The BBC trigger cross section is
23.0$\pm$2.2 mb or ~55\% of the inelastic p+p cross
section. Simulations, and data collected without requiring the BBC
trigger, indicate that the triggered events include 79\% of events
with particles in the central arm acceptance. This number coincides
with the fraction of non-diffractive events triggered by the BBC.  The
bulk of the data sample was recorded requiring a coincidence of the
BBC trigger with a single electron trigger (ERT) that matches hits in
the RICH to 2x2 trigger tiles in the EMCal with a minimum energy of
400 \mevc. In the active area the ERT trigger has a very high
efficiency for electrons; around 500 \mevc\ it reaches approximately
50\% and then saturates around 1 \gevc\ close to 100\%. After applying
an interaction-vertex cut of $\pm$30 cm the total integrated
luminosities were 43 nb$^{-1}$ and 2.25 pb$^{-1}$ for the MB and ERT
trigger, respectively.

\begin{figure}
\includegraphics[width=1.0\linewidth]{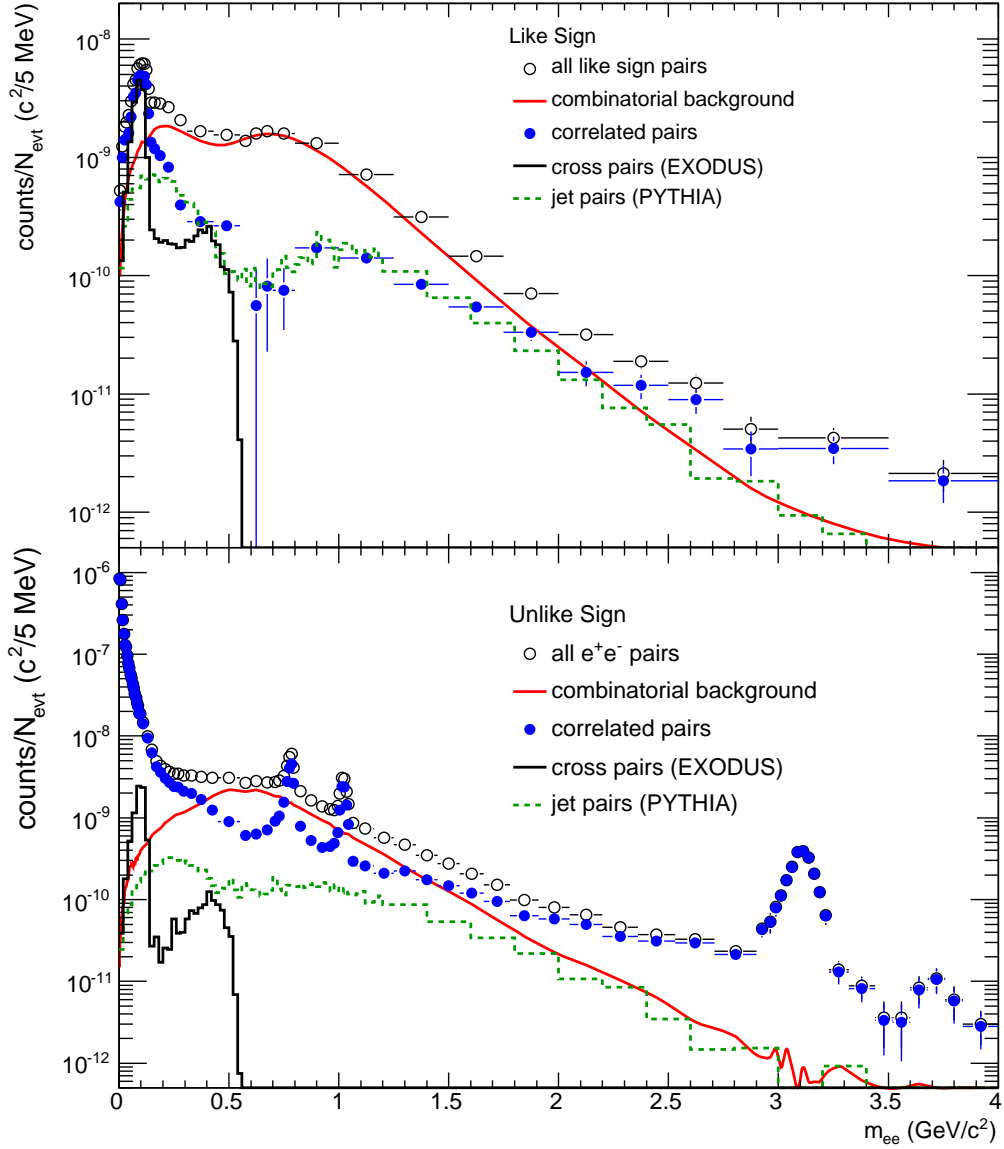}
\caption {\label{fig:rawspectra} Raw dielectron spectra. The top panel
  shows like-sign pairs as measured in the experiment, the
  combinatorial background from mixed events, the correlated pair
  background obtained by subtracting the combinatorial background, and
  the individual contributions from cross and jet pairs to the
  correlated background (see text). The bottom panel shows the same
  distributions for unlike-sign pairs. The correlated background in
  both panels is normalized to the measured like-sign pairs remaining
  after subtracting the combinatorial background.  }
\end{figure}

All electrons and positrons with \pt $>$ 200 \mevc\ are combined into
like- and unlike-sign pairs.  For each pair we check that at least one
of the tracks was registered by the ERT trigger. The event is rejected
if the two tracks of the pair overlap in any of the detectors; this
cut removes 2\% of the \ee pairs. This cut is necessary to assure that
the combinatorial pair background is reproduced from mixed
events. Pairs originating from photon conversions in the detector
material are removed by a cut on the orientation of the pairs in the
magnetic field \cite{Alberica}.  Fig.~\ref{fig:rawspectra} shows the
raw yields as a function of pair mass for both like- and unlike-sign
pairs. The unlike-sign spectrum measures the signal from hadron decays
and open charm plus background, while the like-sign spectrum measures
only the background. Due to the different acceptance for like- and
unlike-sign pairs the shape of the background is different for the two
charge combinations.

We have developed two independent methods to subtract the
background. In the first method we decompose the background into two
components: a combinatorial background made of uncorrelated pairs and
a background of correlated pairs. The combinatorial background is
determined from mixed events using the procedure described in more
detail in \cite{Alberica,ppg075}.  Since our data sample required a
single electron trigger the mixed events are generated from the MB
sample, with the trigger condition applied to each pair, i.e. one of
the tracks must have fired the ERT trigger. The like-sign mixed event
background and the measured like-sign pairs do not have the same
distribution, which is an indication of a correlated background in the
data. However, in the region roughly corresponding to a mass of 300
\mev2, $p_T$ above 400 \mevc\ and a transverse mass $m_T =
\sqrt{m^2+(p_T/c)^2}$ below 1.2 \gev2\ the distributions are very
similar. We therefore normalize the pairs from mixed events to the
data in this region\footnote{The exact region used for the
  normalization is given by the following four conditions $m>$300
  MeV/$c^2$, $m_T<1.2$ MeV/$c^2$, $p_T/c-1.5 m\leq 200$ MeV/$c^2$, and
  $p_T/c-0.75 m\geq 150$ MeV/$c^2$.}. This normalization has a
statistical accuracy of 2.4\%.  After normalization the data show
relatively more yield both at low mass and large $p_T$ as well as at
low $p_T$ and large mass.  By integrating the normalized like-sign
mixed events we determine the number of like-sign background pairs
$N_{++}$ and $N_{--}$, which then give the normalization of the
unlike-sign mixed events as $2\sqrt{N_{++}N_{--}}$.

The mixed event backgrounds as well as the distributions after
subtraction are also shown in Fig.~\ref{fig:rawspectra}. The remaining
pairs, like and unlike, are considered correlated pairs, where the
like-sign distribution only contains correlated background pairs while
the unlike contains also the signal. The correlated background pairs
stem from two sources. ``Cross pairs'' result from decays of single
$\pi^0$ or $\eta$ mesons with two electron pairs in the final state,
such as double Dalitz decays, Dalitz decays plus conversion of the
accompanying photon, and $\gamma\gamma$ decays where both photons
convert. These pairs have a mass lower than the $\eta$ mass of 548
\mev2. Cross pairs were simulated using our hadron decay generator
EXODUS including the PHENIX acceptance \cite{acceptance}.  ``Jet
pairs'' are produced by two independent hadron decays yielding
electron pairs, either within the same jet or in the back-to-back
jets. Jet pairs were simulated using minimum bias events generated
with PYTHIA \cite{pythiamb} with the branching ratio of the $\pi^0$
Dalitz decay set to 100\% to enhance the sample of jet pairs per
event. The resulting \ee pairs are filtered through the PHENIX
acceptance.  Pairs from mixed events are subtracted from the like- and
unlike-sign pair distributions to find the correlated pair
distributions.  This procedure excludes ``signal'' such as unlike-sign
pairs from a single hadron decay. The mixed event background is
normalized by the same method used in the data analysis, described
previously. It was found that correlated pairs from the same jet
typically have small mass and large \pt while those from back-to-back
jets have large mass and smaller \pt. Since the correlated background
pairs populate like- and unlike-sign combinations equally, their yield
was determined by simultaneously fitting simulated cross and jet pair
mass distributions to the measured correlated like-sign pair mass
spectrum. The resulting two normalization factors, one for cross- the
other for jet-pairs, are then applied to the unlike-sign correlated
background. Contributions of both correlated background sources are
also shown in Fig.~\ref{fig:rawspectra}. The signal is extracted by
subtracting the unlike-sign correlated backgrounds from the
distribution of all correlated pairs.

In our second method we make no assumptions about the shape of the
correlated background nor about the decomposition of correlated and
uncorrelated background. The measured like-sign distribution is
corrected for the acceptance difference between like- and unlike-sign
pairs, i.e. the ratio of the acceptance, binned in \pt and mass, of
unlike- to like-sign pairs. Since the acceptance is a function of mass
and $p_T$, we have checked that for different \ee pair sources, which
span reasonable variations in mass and $p_T$ shapes of the \ee pairs,
the relative acceptance is unchanged. The corrected like-sign
distribution is then subtracted from the unlike-sign pairs. Up to 3.5
\gev2\ the difference between the signal extracted using the two
background subtraction techniques agrees to better than
$\pm$10\%. Above 3.5 \gevc\ the difference becomes much larger, which
may indicate additional correlated background.  In this region we
subtract the measured like-sign yield, the larger of our two
background estimates, and include the difference of the two methods as
asymmetric systematic uncertainty on the signal yield.

\begin{figure}
\includegraphics[width=1.0\linewidth]{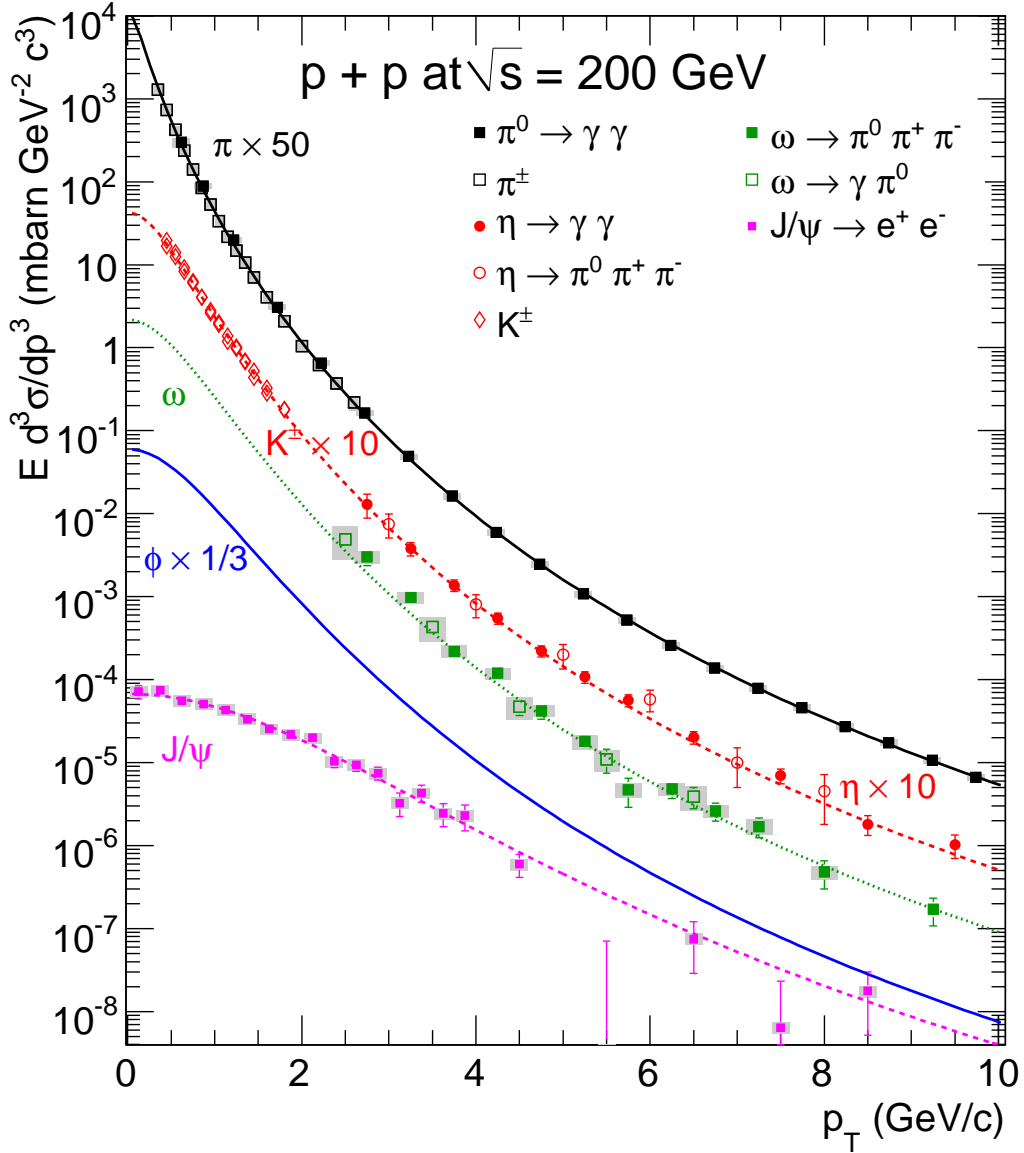}
\caption {\label{fig:mtfits} Compilation of meson production cross
  sections in \pp collisions at \sqrts=200~GeV. Shown are data for
  neutral \cite{pi0} and charged pions \cite{pikp}, $\eta$ \cite{eta},
  Kaons \cite{pikp}, $\omega$ \cite{omega}, and $J/\psi$ \cite{jpsi}.
  The data are compared to the parameterization based on \mt scaling
  used in our hadron decay generator.}
\end{figure}

\begin{figure}
\includegraphics[width=1.0\linewidth]{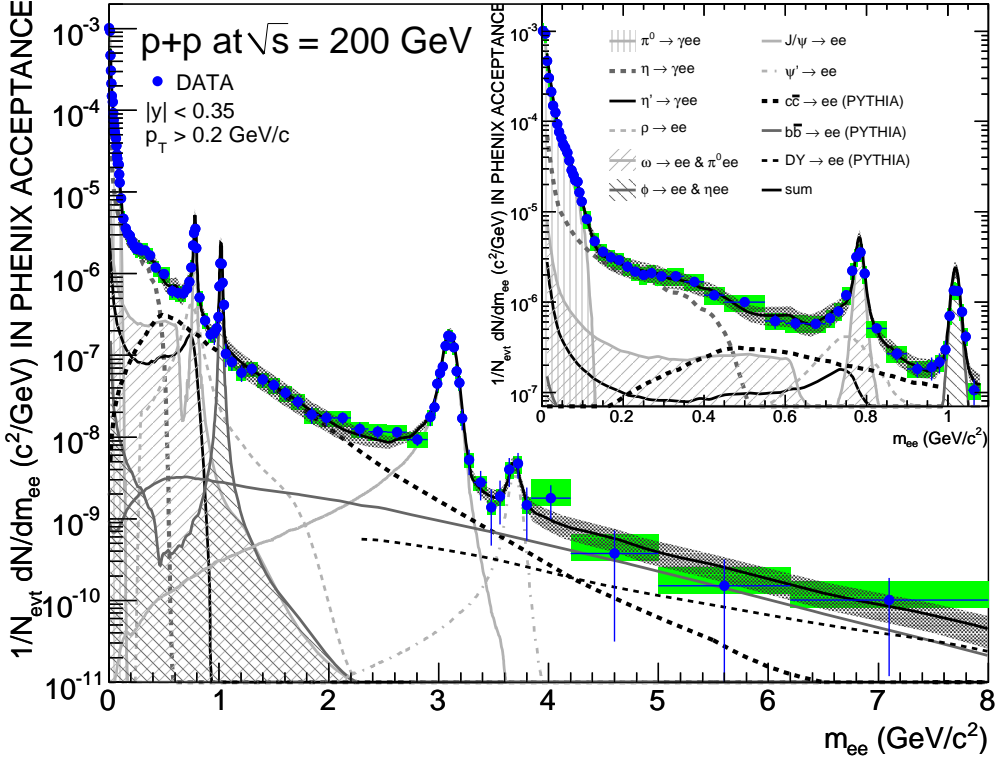}
\caption {\label{fig:massspectrum} Electron-positron pair yield per
  inelastic \pp collision as function of pair mass. Data show
  statistical (bars) and systematic (shades) errors separately. The
  yield per event can be converted to a cross section by multiplying
  with the inelastic \pp cross section of 42.2 mb. The data are
  compared to a cocktail of known sources. The contribution from
  hadron decays is independently normalized based on meson
  measurements in PHENIX, the systematic uncertainties are given by
  the error band. The contribution from open charm production is
  fitted to match the data. The inset shows the same data but focuses
  on the low mass region.  }
\end{figure}

In the next step the signal is corrected for electron reconstruction
efficiency and trigger efficiency. The electron reconstruction
efficiency was determined with a Monte Carlo simulation of the PHENIX
detector (similar to \cite{ppg075}). The trigger efficiency for single
electrons was measured using the MB sample. For each of the 8
calorimeter sectors we determine the ratio of electrons that fired the
ERT trigger to all electrons reconstructed as function of $p_T$. Pairs
from hadron decays simulated with EXODUS are filtered by the
acceptance and then folded with the ERT trigger efficiency to extract
the pair trigger efficiency as function of mass.  At high masses the
trigger efficiency saturates at 72\%, limited by the active area of
the trigger, from 1.5 to 0.5 \gev2\ the pair efficiency gradually
drops to 32\% and remains approximately constant at lower masses. In
addition, the yield is corrected by 0.79/0.55 = 1.44 to account for
the fraction of the inelastic \pp cross section missed by our
interaction trigger.  The systematic uncertainties on the fully
corrected spectrum shown in Fig.~\ref{fig:massspectrum} are summarized in
Table~\ref{tab:errors}.

\begin{table}[th]
  \caption{\label{tab:errors} Systematic uncertainties of the dilepton
    yield due to different sources and for different mass ranges. The
    uncertainties vary with mass and the largest uncertainties are
    quoted for each mass range. The contribution quoted for the jet
    pair subtraction also accounts for the difference in the signal
    observed between our two background subtraction techniques.  }
\begin{center}
\begin{tabular}{c|cccc}
               & $<$0.4 GeV/$c^2$ & 0.4-1.1 GeV/$c^2$ & 1.1-3.5 GeV/$c^2$ & $>$3.5 GeV/$c^2$ \\ 
\hline
minimum bias trigger       &  11.3\%  &   11.3\%  &  11.3\% & 11.3\%  \\
ERT trigger efficiency     &   5\%    &     5\%   & 5\%     &  5\%    \\
conversion rejection       &   5\%    &     -     & -       &  -      \\
mixed event background     &   2\%    &     8\%   & 4\%     &  -      \\
cross pair subtraction     & $<$1\%   &     -     & -       &  -      \\
jet pair subtraction       &   2\%    &     3\%   & 11\%    & +70\%   \\
reconstruction efficiency  & 14.4\%   &  14.4\%   & 14.4\%  & 14.4\%  \\
\hline
total                      & 19.8\%   &  20.8\%   & 22.3\%  & +73\%,$-$19\% \\

\end{tabular}
\end{center}
\end{table}

We model the \ee pair contributions from hadron decays using the
EXODUS decay generator. We follow closely the approach given in
\cite{ppg065,CERES}, however, we have updated all input to match the
most recent PHENIX data.  We assume that all hadrons have a constant
rapidity density in the range $\left|\Delta\eta\right|\le 0.35$ and a
homogeneous distribution in azimuthal angle. Transverse momentum
distributions are based on measurements in the same experiment where
possible. The key input is the rapidity density $dN/dy = 1.06\pm0.11$
of neutral pions, which we determine from a fit to PHENIX data on
charged and neutral pions, as shown in Fig.~\ref{fig:mtfits}. The
functional form of the pion transverse momentum distribution is given
by:

\begin{equation}
\label{eq1}
E\frac{d^3\sigma}{dp3} = A ( e^{-(a p_T + b p_T^2)} + p_T/p_0)^{-n}
\end{equation}

with $A=377\pm60$ mb GeV$^{-2}$c$^3$, $a=0356\pm0.014$
(GeV/$c$)$^{-1}$, $b=0.068\pm0.019$ (GeV/$c$)$^{-2}$, $p_0=0.7\pm0.02$
GeV/$c$ and the power $n=8.25\pm0.04$.  For all other mesons we assume
\mt scaling, replacing \pt by $\sqrt{m^2-m_{\pi}^2+(p_T/c)^2}$, where
$m$ is the mass of the meson. For the $\eta$, $\omega$, $\phi$, and
$J/\psi$ we fit a normalization factor to PHENIX data. In
Fig.~\ref{fig:mtfits} the results are compared to published PHENIX
data; excellent agreement with the data is achieved. The $\eta$ meson
is measured only at higher $p_T$, however, the fit is in good
agreement with the \pt distribution of Kaons, which have similar mass.

In order to extract the meson yield per inelastic \pp collision we
integrate the fits over all $p_T$. Results, systematic uncertainties,
and references to data are given in Table.~\ref{tab:cocktail}. For the
$\rho$ meson we assume $\sigma_\rho/\sigma_\omega=1.15\pm0.15$,
consistent with values found in jet fragmentation \cite{PDG}. The
$\eta'$ yield is scaled to be consistent with jet fragmentation
$\sigma_{\eta'}/\sigma_\eta=0.15\pm0.15$ \cite{PDG}. The $\psi'$ is
adjusted to the value of $\sigma_{\psi'}/\sigma_{J/\psi}=0.14\pm0.03$
\cite{psiprime}. For the $\eta$, $\omega$, $\phi$, and $J/\psi$ the
quoted uncertainties include those on the data as well as those using
different shapes of the \pt distributions to extrapolate to zero
$p_T$. Specifically we have fitted the functional form given in
equation \ref{eq1} with all parameters free and also an exponential
distribution in \mt.  For the $\rho$, $\eta'$, and $\psi'$ the
uncertainty is given by the uncertainty we assumed for the cross
section ratios. We note that the dilepton spectra from meson decays
are rather insensitive to the exact shape of the \pt distribution.

\begin{table}[th]
  \caption{\label{tab:cocktail} Hadron rapidity densities used in our
    hadron decay generator. For the $\omega$ and $\phi$ meson data
    from this analysis were used together with data from the quoted
    references.}
\begin{center}
\begin{tabular}{c|ccc}
               & $\frac{dN}{dy}|_{y=0}$ & relative err. & data used \\ 
\hline
$\pi^0$        & $1.065\pm0.11$     &   10\%        &  PHENIX \cite{pi0},\cite{pikp} \\
$\eta$         & $0.11\pm0.03$      &   30\%        &  PHENIX \cite{eta} \\
$\rho$         & $0.089\pm0.025$    &   28\%        &  jet fragmentation \cite{PDG} \\
$\omega$       & $0.078\pm0.018$    &   23\%        &  PHENIX \cite{omega}      \\
$\phi$         & $0.009\pm0.002$    &   24\%        &  PHENIX \cite{phi}        \\
$\eta'$        & $0.016\pm0.016$    &  100\%        &  jet fragmentation \cite{PDG} \\
$J/\psi$       & $(1.77\pm0.27) \times 10^{-5}$ & 15\%        &  PHENIX \cite{jpsi} \\
$\psi'$        & $(2.5\pm0.7 ) \times 10^{-6}$   &   27\%        & \cite{psiprime} \\

\end{tabular}
\end{center}
\end{table}

Once the meson yields and \pt spectra are known the dilepton spectrum
is given by decay kinematics and branching ratios, which are
implemented in our decay generator EXODUS following earlier work
published in \cite{ppg065,CERES}. The branching ratios are taken from
the compilation of particle properties in \cite{PDG}. For the Dalitz
decays $\pi^0$, $\eta$, $\eta'\rightarrow e^+e^-\gamma$ and the decay
$\omega\rightarrow e^+e^-\pi^0$ we use the Kroll-Wada expression
\cite{kroll-wada} with electromagnetic transition form factors
measured by the Lepton-G collaboration \cite{lepton-g,landsberg}. For
the decays of the vector mesons $\rho$, $\omega$, $\phi\rightarrow
e^+e^-$ we use the expression derived by Gounaris and Sakurai
\cite{gounaris-sakurai}, extending it to 2 \gev2, slightly beyond its
validity range.  For the $J/\psi$ and $\psi'\rightarrow e^+e^-$ we use
the same expression modified to include radiative corrections as
discussed in \cite{jpsi}. The resulting dilepton spectra are compared
to our data in Fig.~\ref{fig:massspectrum} with the systematic
uncertainties shown as a band. They are calculated as a function of
mass and are dominated by the uncertainties on the meson yield
tabulated in Tab.~\ref{tab:cocktail}. The uncertainty from the
measured electromagnetic transition form factors, in particular for
the $\omega\rightarrow\e^+e^-\pi^0$ decay, is also included but
contributes visibly only in the range around 500 to 600 \mev2. Also
shown on Fig.~\ref{fig:massspectrum} are the contributions from open
charm and bottom production, discussed in more detail below, as well
as from the Drell-Yan process, which is negligible.  The data agree
very well with the sum of all known sources.

\begin{figure}
\includegraphics[width=1.0\linewidth]{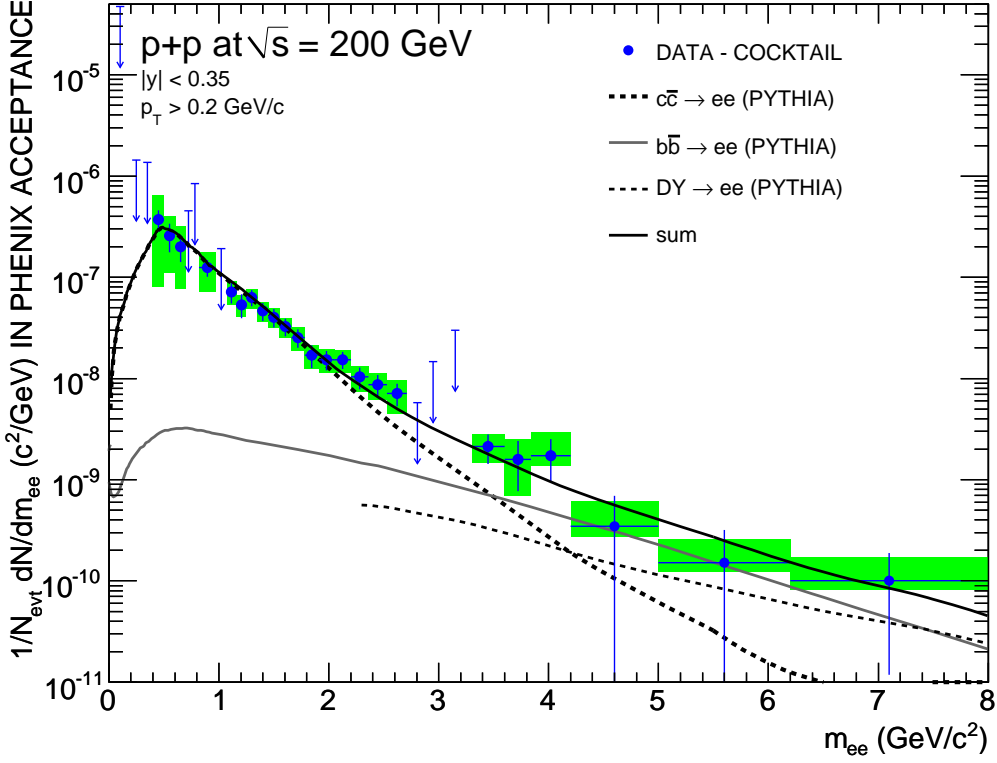}
\caption {\label{fig:charm} Electron-positron mass distributions from
  semileptonic decays of heavy flavor, obtained by subtracting the
  contribution from $\pi^0$, $\eta$, $\omega$, $\rho$, $\phi$,
  $J/\psi$ and $\psi'$ mesons from the inclusive \ee pair yield.  The
  arrows indicate upper limits (95\% CL) in the mass regions where the
  charm contribution is smaller or comparable to the systematic
  uncertainties. For all data points statistical error bars and
  systematic uncertainty boxes, including data and model
  contributions, are shown. Also shown are expected contributions from
  charm, scaled to data, and bottom as well as Drell-Yan. }
\end{figure}

Except for the vector meson peaks, the dilepton yield in the mass
range above 1.1 \gev2\ is dominated by semileptonic decays of D and B
mesons correlated through flavor conservation.  To determine this
contribution we subtract the meson decay cocktail from the dilepton
data, the resulting mass spectrum is shown in Fig.~\ref{fig:charm}. In
the PHENIX acceptance the integrated \ee pair yield per event from
heavy flavor decays in the range from 1.1 to 2.5 \gev2\ is 4.21 $\pm$
0.28(stat) $\pm$ 1.02(syst) $\times 10^{-8}$. The systematic
uncertainties are those tabulated in Tab.~\ref{tab:errors} plus the
uncertainty on the cocktail subtraction.  Since the cocktail
subtraction is dominated by the high mass end of the broad $\rho$
resonance, which is not very well known, we assume 100\% systematic
uncertainty.  To estimate the rapidity density of $c\bar{c}$ pairs the
measured \ee pair yield is corrected for the geometrical acceptance,
i.e. corrected from requiring both electron and positron within the
PHENIX central arm acceptance to having the electron pair within one
unit of rapidity at mid-rapidity.  It then is extrapolated to zero \ee
pair mass and converted to $c\bar{c}$ using known branching ratios of
semileptonic decays \cite{PDG}. This correction is model dependent; we
used our tuned PYTHIA simulation \cite{pythia} to directly relate \ee
pairs from charm in the PHENIX acceptance and in the mass range from
1.1 to 2.5 \gev2\ to the $c\bar{c}$ rapidity density.

For single tracks the acceptance is known to better than 5\%.
Neglecting correlations between the electron and positron this implies
an uncertainty of less than 10\% for pairs. However, the fraction of
\ee pairs from correlated heavy flavor decays at mid-rapidity depends
on the dynamical correlations between the quarks. These are not very
accurately known \cite{e791}, in particular in the azimuthal
direction. Therefore additional systematic uncertainties need to be
considered.  In PYTHIA the intrinsic $k_T$ parameter modifies the
azimuthal correlation between $c$ and $\bar{c}$.  We have varied $k_T$
between 1 and 3 \gev2\ and reevaluate the fraction of \ee pairs at
mid-rapidity. A $\pm$20\% variation was found. Different choices of
parton distribution functions (PDF's) lead to modifications of the
longitudinal correlation of the pair, often expressed as the rapidity
gap between the $c$ and $\bar{c}$ quarks. We used different parton
distribution functions available in PYTHIA, specifically we have used
CTEQ5L, CTEQ4L, GRV94LO, GRV98LO, and MRST(c-g). We find $\pm$11\%
deviations for the \ee pair yield in the PHENIX acceptance. When
converting the \ee pair yield to $c\bar{c}$ pairs there is also a
$\pm$21\% uncertainty resulting from uncertainies of relative
abundance of charmed hadrons and of the branching ratios to
semileptonic decays. We use an effective branching ratio for
$c\rightarrow e$ of 9.5\%$ \pm$ 1\%, which was calculated from
$D^+/D^0 = 0.45 \pm 0.1, D_s/D^0 = 0.25 \pm 0.1$, and $\Lambda_c/D^0 =
0.1 \pm 0.05$ and the branching ratios from \cite{PDG}. The overall
uncertainty on the extrapolation is approximately 33\%.

We also subtract a 7\% contribution from bottom decays and the Drell
Yan mechanism for which we assign a 100\% systematic uncertainty.  For
the bottom cross section we assume 3.7 $\mu$b \cite{jaroschek}, in
agreement with our data above 4 \gev2. Though negligible, we have also
included the contribution from the Drell-Yan mechanism based on a
cross section of 0.04 $\mu$b \cite{DY}.  For the rapidity density of
$c\bar{c}$ pairs at mid-rapidity we find:

\begin{equation}
\frac{d\sigma_{c\bar{c}}}{dy}|_{y=0} = 118.1 \pm 8.4({\rm stat}) \pm 30.7({\rm syst}) \pm 39.5({\rm model}) {\rm\mu b} 
\nonumber
\end{equation}

The systematic uncertainties on the data analysis and on the model
dependent extrapolation are quoted separately.  Using the rapidity
distribution from HVQMNR \cite{hvqmnr} with CTEQ5M \cite{cteq5m} PDF
as in \cite{ppg065}, the total charm cross section is
$\sigma_{c\bar{c}}$ = 544 $\pm$ 39(stat) $\pm$ 142(syst) $\pm$
200(model) $\mu$b.  The extrapolation to $4\pi$ adds another 15\%
systematic uncertainty, which is included in the last term. This
result is compatible with our previous measurement of single electrons,
which gave $\sigma_{c\bar{c}}=$567 $\pm$ 57(stat) $\pm$
224(syst)$\ \mu$b \cite{ppg065}, and with 
the FONLL prediction of
$256^{+400}_{-146}\ \mu$b \cite{fonll}. 
 
Instead of fixing the bottom cross section, we have tried an
alternative approach.  We take the shape of the bottom and charm \ee
pair distributions from PYTHIA filtered into the PHENIX acceptance and
then fit the charm and bottom contribution to the data. For the charm
cross section we obtain $\sigma_{c\bar{c}}=$518 $\pm$ 47(stat) $\pm$
135(syst) $\pm$ 190(model)$\ \mu$b, consistent with our earlier
analysis. The bottom cross section is $\sigma_{b\bar{b}}=$ 3.9 $\pm$
2.5(stat)$^{+3}_{-2}$(syst) $\mu$b.  In addition to the model dependent
systematic uncertainties, which are similar to those on the charm
extraction, the subtraction of \ee pairs from the Drell-Yan mechanism
contributes an extra 10-20\% \cite{vogelsang}.  We estimate that the
combined systematic uncertainty is about 50\% and thus similar to the
statistical error. The value for the bottom cross section is
consistent with our earlier assumption of 3.7 $\mu$b as well as with
the FONLL prediction of $1.87^{+0.99}_{-0.67}\ \mu$b \cite{fonll}.

In conclusion, we have measured \ee pairs in the mass range from 0 to
8 \gev2\ in \pp collisions at \sqrts= 200 GeV. Within the systematic
uncertainties the data can be described by known contributions from
light meson decays, mostly measured in the same experiment, as well as
from semileptonic decays of mesons carrying heavy flavor. The required
charm and bottom production cross sections are consistent with the
upper FONLL predictions and with the PHENIX measurement of single
electrons.

We thank the staff of the Collider-Accelerator and
Physics Departments at BNL for their vital contributions.
We acknowledge support from
the Department of Energy and NSF (U.S.A.),
MEXT and JSPS (Japan),
CNPq and FAPESP (Brazil),
NSFC (China),
IN2P3/CNRS, and CEA (France),
BMBF, DAAD, and AvH (Germany),
OTKA (Hungary), 1
DAE (India),
ISF (Israel),
KRF and KOSEF (Korea),
MES, RAS, and FAAE (Russia),
VR and KAW (Sweden),
U.S. CRDF for the FSU,
US-Hungarian NSF-OTKA-MTA,
and US-Israel BSF.

\def\IJMPA{{Int. J. Mod. Phys.}~{\bf A}}
\def\JPG{{J. Phys}~{\bf G}}
\def\NCA{Nuovo Cimento}
\def\NIM{Nucl. Instrum. Methods}
\def\NIMA{{Nucl. Instrum. Methods}~{\bf A}}
\def\NPA{{Nucl. Phys.}~{\bf A}}
\def\NPB{{Nucl. Phys.}~{\bf B}}
\def\PLB{Phys. Lett. B}
\def\PLC{Phys. Repts.\ }
\def\PRL{Phys. Rev. Lett.\ }
\def\PRD{Phys. Rev. D}
\def\PRC{Phys. Rev. C}
\def\ZPC{{Z. Phys.}~{\bf C}}
\def\EPJ{Eur. Phy. J. {\bf C}}
\def\etal{{\it et al.}}

\bibliographystyle{elsart-num}

\begin{thebibliography}{99}
\bibitem{acceptance} The PHENIX acceptance is parameterized as function of the 
azimuthal angle $\phi$ of a track, its $p_T$, and charge sign q 
by conditions for the DC and the RICH for each spectrometer arm separately: 
$\phi_{min} < \phi + q k_{DC}/p_T < \phi_{max}$ and $\phi_{min} < \phi + q k_{RICH}/p_T < \phi_{max}$
The parameters are $k_{DC}=0.206$ rad \gevc,
$k_{RICH}=0.309$ rad \gevc, $\phi_{min}= -3/16 \pi$ to $\phi_{max}=5/16 \pi$, and
 $\phi_{min}= 11/16 \pi$ to $\phi_{max}=19/16 \pi$.
\bibitem{pythiamb} We used PYTHIA 6.319 changing PYTHIA parameters as follows:
MSEL=0 with the following processes switched on MSUB 11,12,13,28,53,68, 
PARP(91)=1.5 (kt), MSTP(32)=4 (Q2 scale), and CKIN(3)=2.0 (min. parton $p_T$).
\bibitem{pythia} We used PYTHIA 6.205 with CTEQ5L parton distribution
function \cite{cteq5m}. We changed PYTHIA parameters as follows:
PARP(91)=1.5 (kt), PARP(31)=3.5 (K factor), MSTP(33)=1, MSTP(32)=4
(Q2 scale) and in addition for charm production we use MSEL=11 
and PMAS(4,1)=1.25 (mass), for bottom MSEL=5 and PMAS(5,1)=4.1 (mass), 
and for Drell Yan MSEL=11, PARP(31)=1.8, and CKIN(3)=2.0 (min. parton $p_T$).
\end{thebibliography}

\end{document}